\documentclass[10pt,aps,prd,notitlepage,preprintnumbers]{revtex4-1}
\usepackage{graphicx}
\usepackage{slashed}
\usepackage{xcolor}
\usepackage{amsmath}
\usepackage{amssymb}
\usepackage{amsfonts}
\usepackage[pdfborder={0 0 0}]{hyperref}
\newcommand{\ep}{\epsilon}
\newcommand{\la}{\lambda}
\newcommand{\de}{\delta}
\newcommand{\ga}{\gamma}
\newcommand{\al}{\alpha}

\newcommand{\sig}{\sigma}

\newcommand{\no}{\notag\\}
\begin{document}
\title{Transverse momentum dependent gluon fragmentation functions from $J/\psi\ \pi$ production at $e^+ e^-$ colliders}
\author{Guang-Peng \surname{Zhang}}
\affiliation{Center for High Energy Physics, Peking University, Beijing 100871, China}

\begin{abstract}
The back-to-back $J/\psi$ and $\pi$ associated production at $e^+ e^-$ colliders is proposed to detect the gluon transverse momentum dependent(TMD) fragmentation functions. TMD factorization is assumed for this process. With spinless pion, unpolarized and linearly polarized gluon TMD fragmentation functions can be defined. It is found at parton level the hadronic tensor can be described by four structure functions. As a result, there are three independent angular distributions, of which a $\cos{2\phi}$ azimuthal asymmetry is sensitive to the linearly polarized gluon fragmentation function.
\end{abstract}

\maketitle


\section{Introduction}
The transverse momentum dependent fragmentation function(TMDFF) is an important component of transverse momentum dependent(TMD) factorization\cite{Collins:1981uk,Ji:2004wu}. It tells us how the hadron in a jet is affected by the transverse motion and polarizations of the fragmenting parton. Up to now, the quark TMDFFs have been studied very thoroughly(for a review, see\cite{Barone:2010zz}). But for gluon TMDFFs, the study is very limited. In this paper, we point out these functions can be extracted from quarkonium and pion associated production at $e^+e^-$ colliders, i.e. $e^+ + e^-\rightarrow J/\psi+\pi+X$. We will use nonrelativistic QCD(NRQCD)\cite{Bodwin:1994jh} to describe the production of $J/\psi$. In this framework, a pair of heavy quarks is first produced from the hard interaction and then evolves into a quarkonium according to NRQCD. Thus at leading order of $\al_s$, it must be a gluon to fragment into the final pion. If we demand $J/\psi$ and $\pi$ are nearly back-to-back, their relative transverse
momentum distribution will be very sensitive to the transverse motion of the fragmenting gluon, and the cross section should be described by gluon TMDFFs. Since the initial state is colorless, there is no interference between initial and final states and then all soft divergences can be absorbed into the fragmentation function and NRQCD matrix elements and a proper soft factor. In this sense we expect TMD factorization to hold for this process. A potential problem is whether the fragmentation function is general, or is the same as that derived from Semi-inclusive deep inelastic scattering(SIDIS). This can be checked by one-loop calculation. But here we just confine ourselves to the tree level and discuss how much information can be extracted assuming the factorization.
The organization of this paper is as follows: In Sec.II the notations and kinematics are introduced; In Sec.III the formalism and the derivation of three independent angular distributions are given; Sec.IV includes our main result and a discussion; Sec.V is the summary.


\section{Kinematics}
\begin{figure}
\begin{center}
\begin{minipage}{0.4\textwidth}
\includegraphics[scale=0.5]{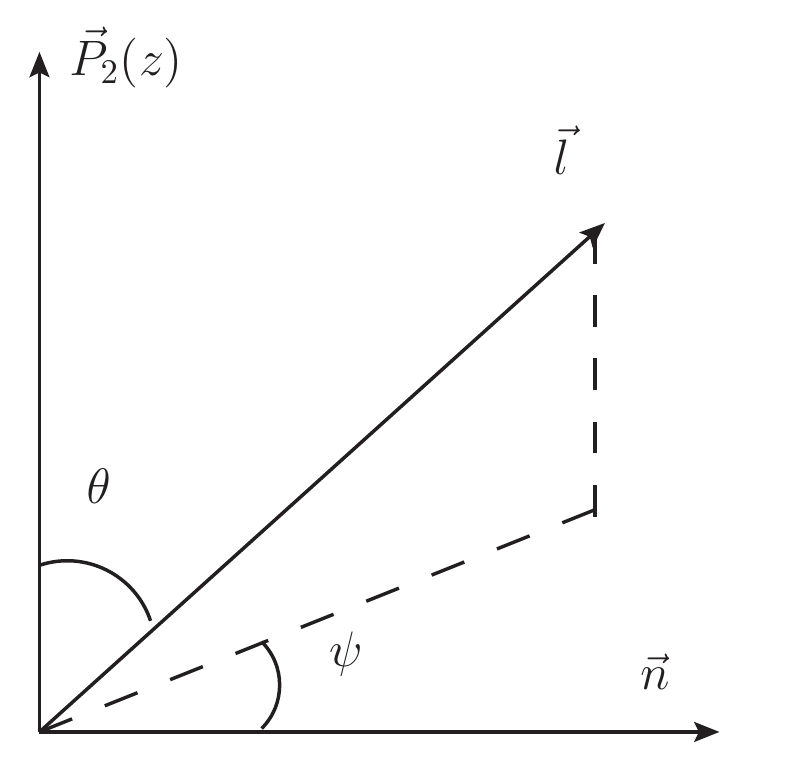}\\
\centering{(a)}
\end{minipage}
\begin{minipage}{0.4\textwidth}
\includegraphics[scale=0.5]{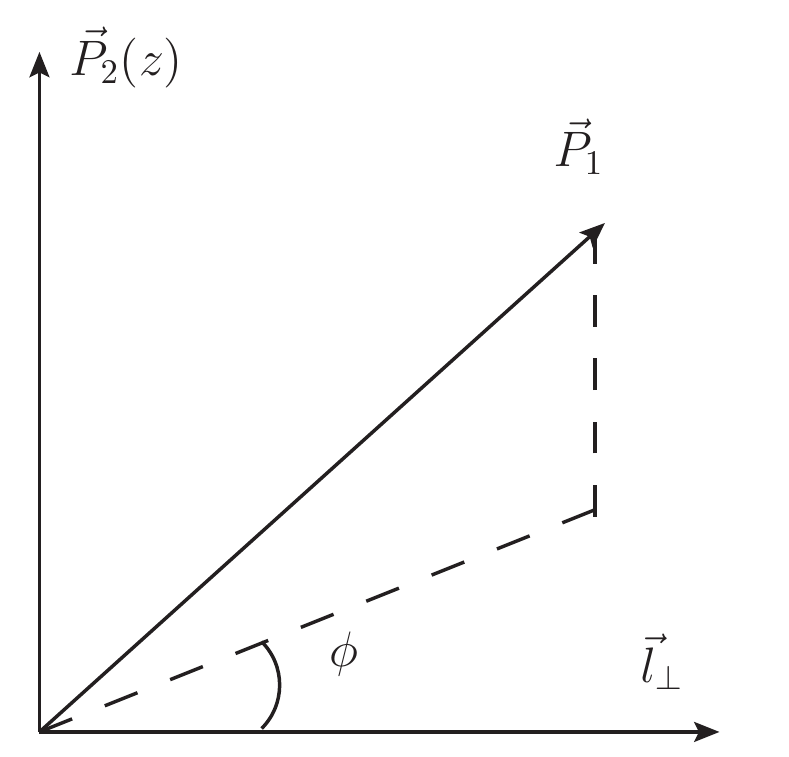}\\
\centering{(b)}
\end{minipage}
\end{center}
\caption{The scattering angles defined in hadron frame, where the momentum of pion $P_2$ is along $+z$-axis. $\vec{l}$ and $\vec{P}_1$ are the momenta of
electron and $J/\psi$, respectively.}\label{fig:frame}
\end{figure}

The process we study is
\begin{align}
e^+(l')+e^-(l)\rightarrow J/\psi(P_1)+\pi(P_2)+X,
\end{align}
where $l',\ l,\ P_1,\ P_2$ are the momenta for each particle, respectively. For this process, it is convenient to work in the hadron frame\cite{Boer:1997qn}. This is a
frame in the center of mass system(CMS) of leptons and with $+z$-axis along the three momentum of pion $\vec{P}_2$. The $+x$-axis can be an arbitrary fixed direction denoted by $\vec{n}$, which is perpendicular to $\vec{P}_2$.
In this frame the momentum of lepton $\vec{l}$ is along the direction $(\theta,\ \psi)$, as shown in Fig.\ref{fig:frame}(a).
In our interested region, $J/\psi$ and $\pi$ are nearly back-to-back, i.e. $\vec{P}^2_{1\perp}\leq \Lambda_{QCD}^2$. Then only the transverse direction of $P_1$ is relevant. The azimuthal angle between $\vec{l}_\perp$ and $\vec{P}_{1\perp}$ is defined as $\phi$, as shown in Fig.\ref{fig:frame}(b). Equivalently, $\phi$ is the angle between the plane expanded by $(\vec{P}_2,\ \vec{l})$ and that by $(\vec{P}_2,\ \vec{P}_1)$.

In our interested region, the invariant mass of leptons $Q^2=q^2=(l+l')^2$ is much higher than the typical nonperturbative scale $\Lambda_{QCD}^2$. In this region, the mass of pion can be ignored, i.e. $P_2^2=0$. Since the quarkonium mass $M_J\simeq 2M_Q$ is also a hard scale, we keep the mass of $J/\psi$ explicit in our calculation, and define
\begin{align}
\tau=\frac{M_J}{Q}=\frac{2M_Q}{Q}.
\end{align}
Now the TMD factorization is expected to be applicable in the region $\vec{P}_{1\perp}^2\ll Q^2$ and $\vec{P}_{1\perp}^2\ll M_J^2$.

The differential cross section can be written as:
\begin{align}
\frac{d\sig}{dz_2 d\Omega dz_1 d^{2}P_{1\perp}}=\frac{e^2 e_Q^2 z_2}{32\pi^2 Q^4\sqrt{z_1^2-\tau^2}}L_{\mu\nu}W^{\mu\nu},
\end{align}
where $\Omega=(\theta,\psi)$ is the solid angle of electron, $e$, $e_Q$ are the electric charges of electron and heavy quark respectively, and $z_1$, $z_2$ are the energy fractions of $J/\psi$ and $\pi$ in CMS frame, that is,
\begin{align}
z_1=\frac{P_1\cdot q}{q^2},\ \ z_2=\frac{P_2\cdot q}{q^2}.
\end{align}
The leptonic and hadronic tensors are the standard ones,
\begin{align}
L_{\mu\nu}=&2(l_\nu l'_\mu+l_\mu l'_\nu-l\cdot l' g_{\mu\nu})-2i\la\ep_{\mu\nu\rho\tau}l^\rho {l'}^\tau,\no
W^{\mu\nu}=&\sum_X\langle 0|j^\nu(0)|J/\psi(P_1)\pi(P_2)X\rangle\langle J/\psi(P_1)\pi(P_2)X|j^\mu(0)|0\rangle\de^4(q-P_1-P_2-P_X),
\end{align}
where $j^\mu=\bar{\psi}\ga^\mu\psi$ is electromagnetic current and $\la$ is the helicity of electron.

For the calculation, it is convenient to define the transverse direction through two light-like vectors: $P_2$ and $\tilde{q}\equiv q-\frac{1}{2z_2}P_2$. The transverse metric and anti-symmetric tensor are defined as:
\begin{align}
&g_\perp^{\mu\nu}=g^{\mu\nu}-n^\mu\bar{n}^\nu-n^\nu\bar{n}^\mu,\ \  \ep_\perp^{\mu\nu}=\ep^{\mu\nu\rho\tau}\bar{n}_\rho n_\tau,\no
&n^\mu\equiv\frac{1}{\sqrt{P_2\cdot\tilde{q}}}\tilde{q}^\mu,\ \ \bar{n}^\mu\equiv\frac{1}{\sqrt{P_2\cdot\tilde{q}}}P_2^\mu,
\end{align}
with $\ep^{0123}=1$. In light-cone coordinates the $+$, $-$ components of a vector $a^\mu$ are $a^+=n\cdot a$, $a^-=\bar{n}\cdot a$, and the transverse component is $a_\perp^\mu=g_\perp^{\mu\nu}a_\nu$.

\section{Formalism and structure functions}
\begin{figure}
\includegraphics[scale=0.5]{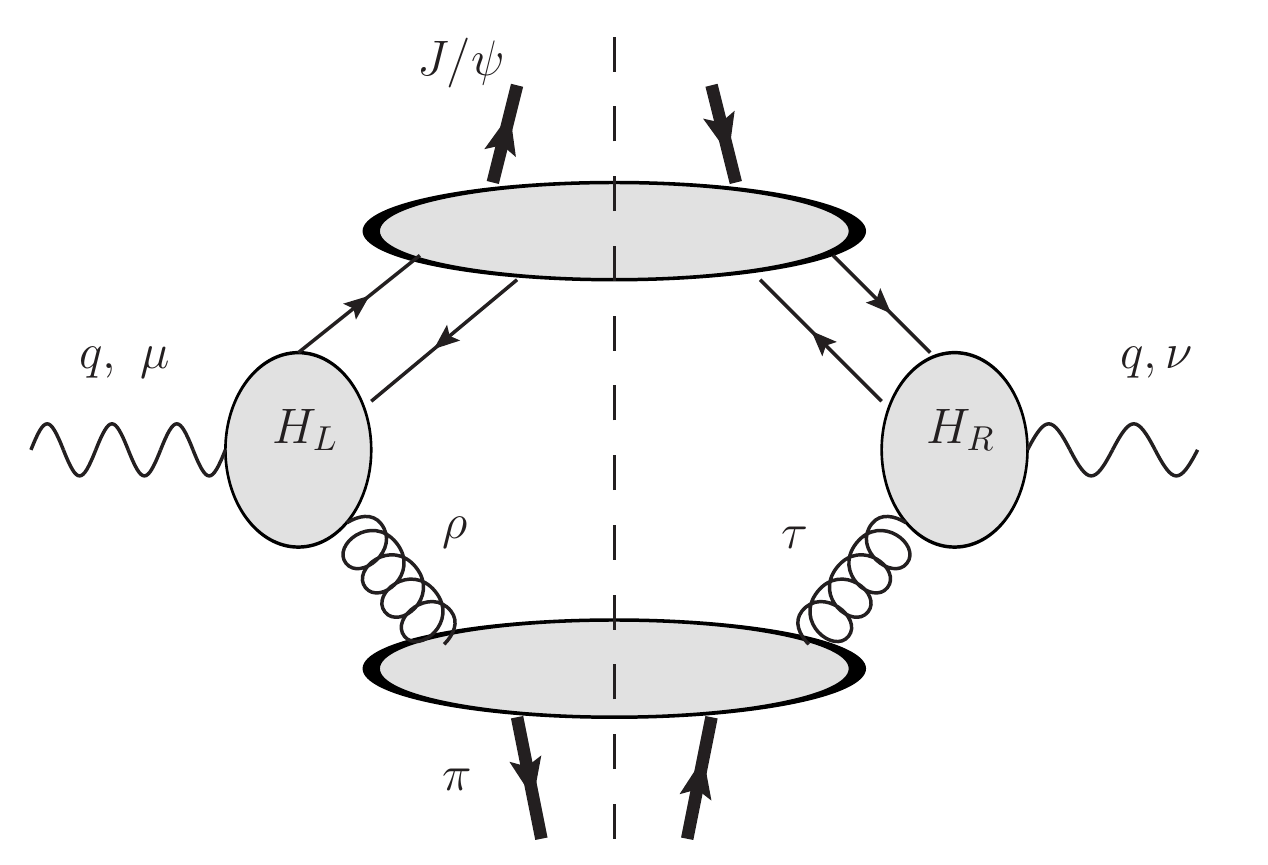}
\caption{The leading region for $J\psi$, $\pi$ back-to-back production. The central bubbles represent the hard scatterings.}\label{fig:factorization}
\end{figure}

At leading power(or twist) of $P_{1\perp}/Q$ and $P_{1\perp}/M_Q$, the cross section can be factorized into the product of a hard kernel and a fragmentation function for the pion
production, as shown in Fig.\ref{fig:factorization}. The fragmentation function is transverse momentum dependent and can be defined as\cite{Mulders:2000sh}:
\begin{align}
&\frac{1}{N_c^2-1}\sum_X\int\frac{d\xi^- d\xi_\perp^2}{(2\pi)^3}e^{ik^+\xi^-+ik_\perp\cdot\xi_\perp}
\langle 0|G_{\perp a}^{+\tau}(\xi^-,\xi_\perp)|P_h X\rangle\langle P_h X|G_{\perp a}^{+\rho}(0)|0\rangle|_{\xi^+=0}\no
=&\frac{P_h^+}{M_h}[-g_\perp^{\rho\tau}\hat{G}(z,k_\perp^2)
+\frac{1}{2M_h^2}(2k_\perp^\rho k_\perp^\tau-g_\perp^{\rho\tau}k_\perp\cdot k_\perp)\hat{H}(z,k_\perp^2)],
\end{align}
where $G_{\perp a}^{\mu\rho}$ is the gluon field strength tensor, $a$ is the color index. Here we have assumed the hadron $h$ is moving along $+z$-axis, then the large component of its momentum is $P_h^+$. The parton momentum fraction is $1/z=k^+/P_h^+$, $0\leq z\leq 1$. We have suppressed the gauge links, which are irrelevant to our discussions here. From the definition, $\hat{G}$ corresponds to the unpolarized fragmenting gluon and $\hat{H}$ corresponds to the linearly polarized gluon. Since the final hadron is spinless or unpolarized, there are only these two fragmentation functions at leading twist.

For the quarkonium production, we use NRQCD\cite{Bodwin:1994jh}. The heavy quark pair $Q\bar{Q}$ is first produced from the hard interaction, then transforms into a quarkonium.
The $Q\bar{Q}$ has a small relative velocity $v$ in the rest frame of the quark pair. NRQCD gives a consistent power expansion according to $v$. To a certain
power of $v$, only finite number of NRQCD matrix elements contribute. For the process considering now, the quark pair from hard interaction must be in color octet. To leading order of $v$, there are three types of NRQCD operators contributing to $J/\psi$ production: $\mathcal{O}_8(^1S_0)$, $\mathcal{O}_8(^3S_1)$,
$\mathcal{O}_8(^3P_J)$\cite{Cho:1995vh,Cho:1995ce}. The quantum number $^{2S+1}L_J$ represents the angular momentums of the quark pair in the rest frame. Here we use the covariant formalism in \cite{Petrelli:1997ge} to project the quark pair to definite partial waves. For details please see \cite{Petrelli:1997ge}, here we just list the final results.

According to the collinear power counting rule\cite{Collins:2008sg,Ma:2013aca}, at leading twist the fragmenting gluon is collinear to the final pion, i.e. $k_g^\mu=(k_g^+,k_g^-,k^\mu_{g\perp})\simeq Q(1,\la^2,\la)$ with $\la\simeq\Lambda_{QCD}/Q\ll 1$. This results in the leading region shown in Fig.\ref{fig:factorization}. Contrary to the collinear factorization, now the transverse momentum $k_{g\perp}$ cannot be ignored in the hard part. Actually, this momentum is contained in the delta function for momentum conservation, i.e. $\de^2(P_{1\perp}+k_{g\perp})$. Since this delta function has already been of leading power, $k_{g\perp}$ can be ignored in other parts of the hard scattering amplitude. Following the procedure in e.g.\cite{Ma:2013aca}, the resulted hadronic tensor is
\begin{align}
W^{\mu\nu}=\frac{2z_2}{(1-\tau^2)Q^2 M_\pi}\de(z_1-z^*_1)\int d^2k_\perp \de^{(2)}(P_{1\perp}+k_{\perp})\mathcal{M}^{\mu\nu}_{\rho\tau}
[-g_\perp^{\rho\tau}\hat{G}(z,k_\perp^2)+\frac{2k_\perp^\rho k_\perp^\tau-g_\perp^{\rho\tau}k_\perp\cdot k_\perp}{2M_\pi^2}\hat{H}(z,k_\perp^2)].\label{hadron_tensor}
\end{align}
where $\mu\nu\rho\tau$ are Lorentz indices, as shown in Fig.\ref{fig:factorization}. Notice that in this formula the energy fraction of $J/\psi$ $z_1$ is totally fixed to be $z_1^*=(1+\tau^2)/2$. This is a result of $P_{1\perp}\rightarrow 0$. Moreover, the momentum fraction $z$ is determined by $z_1$ and $z_2$ definitely. With $z_1=z_1^*$, $z$ and $z_2$ satisfy a very simple relation:
\begin{align}
\frac{1}{z}=\frac{1-\tau^2}{2z_2}.\label{eq:z-z_2}
\end{align}
The tensor $\mathcal{M}^{\mu\nu}_{\rho\tau}$ is the module of the hard amplitude,
\begin{align}
\mathcal{M}^{\mu\nu}_{\rho\tau}=\sum_{LSJ} \frac{1}{N_J N_{color}}\mathcal{A}^\mu_\rho(\mathcal{A}^\nu_\tau)^*\langle\mathcal{O}_8(^{2S+1}L_J)\rangle.
\end{align}
where we have summed over all possible partial waves with $S\leq 1$ and $L\leq 1$; $\mathcal{A}^\mu_\rho$ is the hard partial wave amplitude, but without the polarization vectors for fragmentation gluons.
The average over the color and polarizations of the quark pair is necessary for production processes, since the color and polarizations have been summed over in the NRQCD matrix elements\cite{Petrelli:1997ge}. As the quark pair is in color octet, $N_{color}=N_c^2-1$. For different partial waves $J=0,1,2$, $N_J=1,3,5$
respectively.

As we stated before, $k_\perp$ has been set to zero in $\mathcal{A}^\mu_\rho$, thus $\mathcal{M}^{\mu\nu}_{\rho\tau}$ is independent of $k_\perp$.
In eq.(\ref{hadron_tensor}), the integration over $k_\perp$ can be done independently. That is,
\begin{align}
&\int d^2k_\perp \de^{(2)}(P_{1\perp}+k_{\perp})[-g_\perp^{\rho\tau}\hat{G}(z,k_\perp^2)+\frac{2k_\perp^\rho k_\perp^\tau-g_\perp^{\rho\tau}k_\perp\cdot k_\perp}{2M_\pi^2}\hat{H}(z,k_\perp^2)]\no
=&-g_\perp^{\rho\tau}\mathcal{C}[w_G(k_\perp,h_\perp)\hat{G}(z,k_\perp^2)]-(g_\perp^{\rho\tau}+2h_\perp^\rho h_\perp^\tau)
\mathcal{C}[w_H(k_\perp, h_\perp)\hat{H}(z,k_\perp^2)],\no
&\mathcal{C}[w(k_\perp, h_\perp)f(z,k_\perp^2)]\equiv\int d^2k_\perp \de^{(2)}(k_\perp+P_{1\perp})w(k_\perp, h_\perp)f(z,k_\perp^2),
\label{integration_k_perp}
\end{align}
where $h_\perp^\mu\equiv P_{1\perp}^\mu/\sqrt{\vec{P}_{1\perp}^2}$ satisfies $h_\perp^2=-1$ and
\begin{align}
w_G(k_\perp,h_\perp)=1,\ \ w_H(k_\perp, h_\perp)=\frac{-2(\vec{k}_\perp\cdot\vec{h}_\perp)^2+\vec{k}_\perp^2}{2M_\pi^2}.
\end{align}

Substituting eq.(\ref{integration_k_perp}) into eq.(\ref{hadron_tensor}), one can see the hard part just depends on two tensors
\begin{align}
\mathcal{M}_G^{\mu\nu}=-g_\perp^{\rho\tau}\mathcal{M}^{\mu\nu}_{\rho\tau},\ \
\mathcal{M}_H^{\mu\nu}=-(g_\perp^{\rho\tau}+2h_\perp^\rho h_\perp^\tau)\mathcal{M}^{\mu\nu}_{\rho\tau}.
\end{align}
From this structure, the two tensors are just two different gluon helicity amplitudes. Further simplification can be obtained by decomposing the tensor into
structure functions. This is very easy to realize, because the independent momenta are just $q$, $P_2$, $h_\perp$, and only $h_\perp$ is transverse. Moreover,
$\ga_5$ in the partial wave projections always appears in pair, so there will be no $\ep$-tensor in $\mathcal{M}^{\mu\nu}_{\rho\tau}$ and $\mathcal{M}_{G,H}^{\mu\nu}$. To make the expressions simpler, we put $\mathcal{M}_{G}^{\mu\nu}$ and $\mathcal{M}_{H}^{\mu\nu}$ together to form a two dimensional vector, i.e., $\mathcal{M}^{\mu\nu}=\{\mathcal{M}_{G}^{\mu\nu},\ \mathcal{M}_{H}^{\mu\nu}\}$. The same rule also applies to $F_i$ and $D_i$ defined later.  The decomposition leads to six independent structure functions as follows:
\begin{align}
\mathcal{M}^{\mu\nu}=&F_1(g^{\mu\nu}-\frac{q^\mu q^\nu}{q^2})+F_2\frac{\tilde{P}_2^\mu \tilde{P}_2^\nu}{\tilde{P}_2^2}
+F_3 h_\perp^\mu h_\perp^\nu+F_4 \tilde{h}_\perp^\mu \tilde{h}_\perp^\nu\no
&+F_5(\tilde{P}_2^\mu q^\nu+\tilde{P}_2^\nu q^\mu)+F_6(\tilde{P}_2^\mu q^\nu-\tilde{P}_2^\nu q^\mu),
\end{align}
where $\tilde{P}_2= P_2-z_2 q$ and $\tilde{h}^\mu_\perp=\ep_\perp^{\mu\nu}h_{\perp\nu}$, which satisfy
\begin{align}
\tilde{P}_2\cdot q=0,\ \ g_\perp^{\mu\nu}=-h_\perp^\mu h_\perp^\nu-\tilde{h}^\mu_\perp \tilde{h}^\nu_\perp.
\end{align}
From QED gauge invariance $q_\mu \mathcal{M}^{\mu\nu}=q_\nu \mathcal{M}^{\mu\nu}=0$, one has $F_5=F_6=0$, which is confirmed by our calculation. Thus only four structure functions are nontrivial.
It is clear that $F_i$'s do not depend on the scattering angles $(\theta,\psi,\phi)$. Thus all types of angular distribution can be obtained by contracting
$\mathcal{M}^{\mu\nu}$ with leptonic tensor $L^{\mu\nu}$. The resulted cross section is
\begin{align}
\frac{d\sig}{dz_2 d\Omega dz_1 d^{2}P_{1\perp}}=\left(\frac{2}{3}\right)^2\frac{2\al_{em}^2}{Q^4}\frac{z_2^2 \de(z_1-z_1^*)}{(1-\tau^2)^2 M_\pi}
\sum_{K=G,H}\mathcal{C}[w_K f_K]\left(D^K_1-\frac{1}{2}\sin^2\theta D^K_3+\frac{1}{2}\sin^2\theta\cos 2\phi D^K_2 \right),
\end{align}
with $f_G=\hat{G}(z,k_\perp^2)$, $f_H=\hat{H}(z,k_\perp^2)$ and $w_{G,H}$ their corresponding weights. So there are only three independent angular distributions for this process. Especially, there is a $\cos 2\phi$ azimuthal asymmetry, which will be an important signal for linear fragmentation gluon.
The coefficients $D_i$ are the superpositions of $F_i$,
\begin{align}
D_1=F_3+F_4-2F_1,\ \ D_2=F_4-F_3,\ \ D_3=F_3+F_4+2F_2.
\end{align}
NRQCD matrix elements are contained in these coefficients. To see the contribution of different partial waves, we write out the matrix elements explicitly,
\begin{align}
D_i=D_i(^1S_0)\langle\mathcal{O}_8(^1S_0)\rangle+D_i(^3S_1)\langle\mathcal{O}_8(^3S_1)\rangle+\frac{1}{3}D_i(^1P_1)\langle\mathcal{O}_8(^1P_1)\rangle+\sum_{J=0}^2 D_i(^3P_J)\frac{1}{N_J}\langle\mathcal{O}_8(^3P_J)\rangle,\ i=1,2,3.
\end{align}

\section{Result}
\begin{figure}
\includegraphics[scale=0.5]{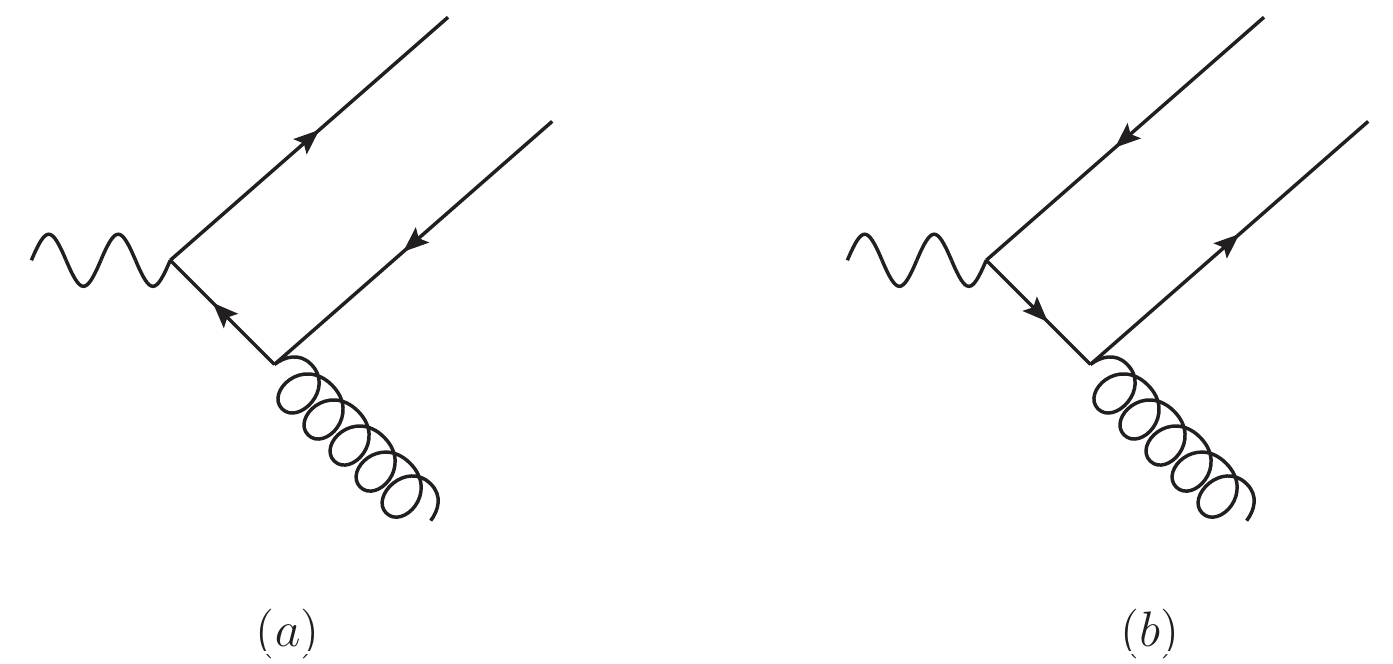}
\caption{The two Feynman diagrams for the hard subprocess at tree level.}\label{fig:Feynman}
\end{figure}

As we can see, the analysis in last section is independent of the details of Feynman diagrams. Since in TMD factorization the hard coefficients just receive contributions from the virtual correction\cite{Ji:2004wu}, the formalism also applies to higher order corrections in $\al_s$. At tree level the calculation is very simple. There are only two Feynman diagrams, as shown in Fig.\ref{fig:Feynman}.

After the calculation, we find that the partial waves $^3S_1$ and $^1P_1$ have no contribution to all $D_i$'s. The linearly polarized gluon fragmentation function only contributes to the $\cos 2\phi$ angular distribution, while unpolarized gluon fragmentation function does not contribute to this angular distribution at all, that is,
\begin{align}
D_2^G=0,\ \ D_1^H=D_3^H=0.
\end{align}
These two equations hold for all partial waves. Based on these facts, our final cross section can be written as
\begin{align}
\frac{d\sig}{dz_2 d\cos\theta d^{2}P_{1\perp}}=&\left(\frac{2}{3}\right)^2\frac{4\pi\al_{em}^2}{Q^4}\frac{z_2^2}{(1-\tau^2)^2 M_\pi}
\left\{\mathcal{C}[w_G \hat{G}]\left(D^G_1-\frac{1}{2}\sin^2\theta D^G_3\right)
+\mathcal{C}[w_H \hat{H}]D^H_2\left(\frac{1}{2}\sin^2\theta\cos 2\phi\right)\right\},\label{eq:main_result}
\end{align}
and
\begin{align}
\left[g_s^2\frac{16(N_c^2-1)^2}{3M_Q^3(1-\tau^2)^2}\right]^{-1}D^G_{1,3}=&
\tilde{D}^G_{1,3}(^1S_0)\langle\mathcal{O}_8(^1S_0)\rangle+\sum_{J=0}^2\tilde{D}^G_{1,3}(^3P_J)\frac{1}{N_J}\langle\mathcal{O}_8(^3P_J)\rangle,\no
\left[g_s^2\frac{16(N_c^2-1)^2}{3M_Q^3(1-\tau^2)^2}\right]^{-1}D^H_2=&
\tilde{D}^H_2(^1S_0)\langle\mathcal{O}_8(^1S_0)\rangle+\sum_{J=0}^2\tilde{D}^H_2(^3P_J)\frac{1}{N_J}\langle\mathcal{O}_8(^3P_J)\rangle.
\end{align}
This is our main result. The corresponding coefficients $\tilde{D}_i$ can be found in Table.\ref{Tab:hard-coef}. Notice that in eq.(\ref{eq:main_result})
$z_1$ and $\psi$ have been integrated over since $z_1$ is just contained in a delta function and the hard coefficients are $\psi$ independent. In experiment this cross section is easier to measure.
Now it is clear that the gluon TMD fragmentation functions can be extracted from the three independent angular distributions. By fitting the experiment data, the constraint on the involved four color octet matrix elements can also be obtained, which are not determined very well in literature( see \cite{Zhang:2009ym} and references therein). Two features of our tree level result should be stressed here. One is about the $z_2$ dependence of the hard coefficient. As can be seen from Table.\ref{Tab:hard-coef}, these hard coefficients are $z_2$ independent. This feature
will remain to higher order corrections, because the hard subprocess just depends on $P_1$ and $q$, which are $z_2$ independent. Another feature is the
threshold enhancement which appears when $\tau^2\rightarrow 1$. This can be easily understood. When the fragmenting gluon becomes soft, the intermediate heavy quark propagator, as shown in Fig.\ref{fig:Feynman}, is approaching to the mass shell, and this causes a factor $1/P_1^z$. Similar enhancement also appears in phase space integration, i.e., $dP_1^z\propto Q^2dz_1/P_1^z$. Since near threshold $P_1^z$ is a small quantity proportional to $\sqrt{1-\tau^2}$, this factor results in an enhancement. This feature
will also remain to higher order corrections.
Since our knowledge about gluon TMDFFs and the four involved NRQCD matrix elements is very limited, here we cease to give a numerical estimate to the cross section.
\begin{table}
\begin{tabular}{|c|cccc|}
\hline
                & $^1S_0$                & $^3P_0$           & $^3P_1$       & $^3P_2$\\
\hline
$\tilde{D}^G_1(^{2S+1}L_J)$ &\ \ $3Q^2\tau^2(\tau^2-1)/8$\ \ & \ \ $(1-3\tau^2)^2/2$\ \  &\ \ $3$\ \  & \ \ $6\tau^4+1$\ \ \\

$\tilde{D}^G_3(^{2S+1}L_J)$ &\ \ $3Q^2\tau^2(\tau^2-1)/8$\ \ & \ \ $(1-3\tau^2)^2/2$\ \  &\ \ $-3(2\tau^2-1)$\ \  & \ \ $6\tau^4-6\tau^2+1$\ \ \\

$\tilde{D}^H_2(^{2S+1}L_J)$ &\ \ $-3Q^2\tau^2(\tau^2-1)/8$\ \ &\ \ $(1-3\tau^2)^2/2$\ \  &\ \ $-3$\ \  &\ \  $1$\ \ \\
\hline
\end{tabular}
\caption{The hard coefficients for different partial waves.}\label{Tab:hard-coef}
\end{table}

Before extracting these nonperturbative quantities from experiments, the most urgent task is to give a clear examination for the TMD factorization for this
process at least to one-loop level, since the kinematics is so simple. At one-loop level, two types of divergence will appear. One
is caused by the collinear gluon connected to the fragmenting gluon, the other is caused by the soft gluon connected to the heavy quark. For the former, collinear power counting works, the divergence will be absorbed into the fragmentation functions by using Ward identities since all external particles of the hard subprocess are on-shell.
For the latter, the situation is a little more complicated. There are two cases. First, an octet heavy quark pair from the hard interaction may emit a real soft gluon to transmit to a color singlet quark pair. Up to the power of $v$ we considered here, the singlet can only be $Q\bar{Q}(^3S_1)$, i.e. an S-wave singlet. In this case the soft divergence is cancelled out after one sums up the divergences from heavy quark and antiquark\cite{Bodwin:1994jh}. Thus color singlet will not affect the factorization even at one-loop level. Second, after emitting a soft gluon, the color octet heavy quark pair may be still in a color octet. In this case, the soft divergences cannot be cancelled out by summing up all relevant diagrams, since we have detected the momenta of final particles. So an additional soft factor is
required. Besides the soft divergences, to higher order of $\al_s$ the $q\bar{q}$ channel, i.e.,$\ga^*\rightarrow q\bar{q}\rightarrow q\bar{q}+Q\bar{Q}$ may break TMD factorization. The heavy quark pair is generated by a fragmenting gluon, and the final state pion is generated by a fragmenting quark. In this case,
the fragmenting gluon has a off-shellness of $M_J^2\simeq 4M_Q^2$, then the contribution from this channel will be suppressed by $P_{1\perp}^2/M_J^2$, and should be ignored since both $M_J^2$ and $Q^2$
are hard scales of the same order in our interested kinematical region. On the other hand, the gluon TMD fragmentation functions may also appear in
some SIDIS processes, such as $e+P\rightarrow A+B+X$\cite{Boer:2011fh,Boer:2010zf,Zhu:2013yxa} where $A, B$ are two hadrons almost back-to-back. It is interesting to see whether the TMD fragmentation functions there are the same as these we used here. The detailed discussion of these issues is beyond the scope of this paper and will be put into a future paper.

\section{Summary}
In this paper we analyzed the back-to-back $J/\psi\ \pi$ associated production at $e^+e^-$ colliders, making use of TMD factorization and NRQCD. The hadron frame where the final pion is along $+z$-axis is convenient for the analysis. In this frame, we find three independent angular distributions which are sensitive
to the unpolarized and linearly polarized gluon TMD fragmentation functions. Especially, the linearly polarized gluon fragmentation function
will contribute to the azimuthal asymmetry. At tree level, NRQCD matrix elements $\langle\mathcal{O}_8(^1S_0)\rangle$ and $\langle\mathcal{O}_8(^3P_J)\rangle$ contribute to these angular distributions. Their information can also be extracted by fitting the data at $e^+e^-$ colliders.

\begin{acknowledgements}
The author would like to thank Dr. G.Y.Tang for useful discussions about the mathematica procedure.
\end{acknowledgements}

\bibliography{cf/ref2}

\end{document}